Comments on "Thermodiffusion: The physico-chemical mechanics view. J. Chem. Phys. 154, 024112 (2021)"

Zi-Kui Liu

Department of Materials Science and Engineering, The Pennsylvania State University

In a series of publications[1–3], Kocherginsky and Gruebele presented a systematic framework for chemical transport and thermodiffusion to predict the Soret coefficients from thermodynamics. A macroscopic derivation of the Onsager reciprocal relations without recourse to microscopic fluctuations or equations of motion was also discussed.

Based on the continuity equation for diffusive transport and the Smoluchowski equation, they defined the flux of a species $i$ as follows (Eq. 3 in ref. [3])

$$J_i = -U_i C_i \frac{\partial \mu_{gi}}{\partial x} \quad \text{Eq. 1}$$

where $U_i$, $C_i$ and $\mu_{gi}$ are Einstein's mobility, concentration, and physicochemical potential of species $i$, respectively, and $x$ is the position in the system. They emphasized that $\mu_{gi}$ is defined in terms of two factors driving the mass transport: an externally applied force and a concentration gradient, and further expressed it using the following equation

$$\mu_{gi}(x) = \mu_i(x) + \mu_{fi}(x) = \mu'_{0i}(T,P) + RT \ln C_i(x) + \mu_{fi}(x) \quad \text{Eq. 2}$$

where $\mu_i(x)$ is the chemical potential at the temperature, pressure, and position in the sample where the measurement is carried out, $\mu_{fi}(x)$ is the potential due to all external forces, such as applied electric fields, $\mu'_{0i}$ includes the activity coefficient and depends on the position via thermodynamic variables such as $T$ and $P$, and $RT \ln C_i(x)$ denotes the contribution from an ideal solution.

The gradient of $\mu_{gi}(x)$ was further derived as follows, i.e. Eq. 9 in ref. [3]

$$\frac{\partial \mu_{gi}}{\partial x} = \frac{\partial \mu'_{0i}}{\partial T}\frac{\partial T}{\partial x} + \frac{\partial \mu'_{0i}}{\partial P}\frac{\partial P}{\partial x} + \frac{\partial \mu'_{0i}}{\partial x} + \frac{\partial \mu_{fi}}{\partial x} + RT \frac{\partial (\ln C_i)}{\partial x} \quad \text{Eq. 3}$$

The significance of Eq. 1 is very important and needs to be further emphasized here. This is different from the commonly used general phenomenological relation, such as Eq. 1 in ref. [1], as follows

$$J_m = \sum_n L_{mn} F_n \quad \text{Eq. 4}$$

Eq. 4 is the same as Eq. 1 in the present paper if $L_{mn} = 0$ for $m \neq n$, i.e. the $L_{mn}$ matrix is diagonal, which naturally fulfils the Onsager reciprocal relations of $L_{mn} = L_{nm}$. This was also shown in our recent work on prediction of Seebeck coefficients for thermoelectric materials [4,5] and in general discussion of cross-phenomena based on the combined law of thermodynamics with the entropy production term kept [6].

However, there are some confusions about Eq. 2 and Eq. 3 above. One is reminded that $\mu_{gi}$ is an internal state function of independent internal state variables of a state which can be stable, metastable, or unstable [6–8]. The distributions of internal state variables inside a system depend on the constraints imposed from the surroundings and the properties of the system, while the position dependence of $\mu_{gi}$ is related to the position dependences of internal state variables and probably their gradients too (see further discussions below). It is important to point out the position itself is not an internal state variable. The internal state variables include temperature, pressure, concentration, and other potentials which could be gravitational and electric fields as mentioned by Kocherginsky and Gruebele [1–3], and they are all position-dependent in the system, i.e. $T(x)$, $P(x)$, $C_i(x)$, and $\Phi_k(x)$.

The chemical potential of species $i$, or physicochemical potential as defined by Kocherginsky and Gruebele, can thus be written as

$$\mu_{gi}(x) = \mu_{gi}(T, P, C_j, \Phi_k) = \mu_{0i}(T, P, \Phi_k) + RT\ln a_i(T, P, C_j, \Phi_k) \quad \text{Eq. 5}$$
$$= \mu_{0i}(T, P, \Phi_k) + RT\ln f_i(T, P, C_j, \Phi_k) + RT\ln C_i(x)$$
$$= \mu'_{0i}(T, P, C_j, \Phi_k) + RT\ln C_i(x)$$

where the subscript $j$ represents all species in the system, $a_i(T, P, C_j, \Phi_k)$ and $f_i(T, P, C_j, \Phi_k)$ are the chemical activity and chemical activity coefficient of species $i$, and $\mu'_{0i}(T, P, C_j, \Phi_k)$ is the same quantity defined in Eq. 2 and depends on concentrations of all species and other fields in addition to temperature and pressure. It is thus evident that introducing $\mu'_{0i}(T, P, C_j, \Phi_k)$ does not simply further derivations, and the assumption by Kocherginsky and Gruebele [1–3] that $\mu'_{0i}$ is only a function of $T$ and $P$ as written in Eq. 2 is not justified though it is also considered as a function of position as shown in Eq. 3, which is the confusion as mentioned above.

Consequently, the gradient of $\mu_{gi}$ can be written as

$$\frac{d\mu_{gi}}{dx} = \frac{\partial \mu_{gi}}{\partial T}\frac{dT}{dx} + \frac{\partial \mu_{gi}}{\partial P}\frac{dP}{dx} + \sum_j \frac{\partial \mu_{gi}}{\partial C_j}\frac{dC_j}{dx} + \sum_k \frac{\partial \mu_{gi}}{\partial \Phi_k}\frac{d\Phi_k}{dx}$$
$$= -S_i \frac{dT}{dx} + V_i \frac{dP}{dx} + \sum_j \phi_{ij} \frac{dC_j}{dx} + \sum_k \alpha_{ik} \frac{d\Phi_k}{dx} \quad \text{Eq. 6}$$

where $S_i$ and $V_i$ are the partial molar entropy and partial molar volume of species $i$, $\phi_{ij} = \frac{\partial \mu_{gi}}{\partial C_j}$ is the thermodynamic factor between species $i$ and $j$, and $\alpha_{ik} = \frac{\partial \mu_{gi}}{\partial \Phi_k}$ is the derivative of chemical potential of species $i$ to the potential $k$. The Soret coefficient of species $i$ is thus simply its partial molar entropy, also mentioned by Duhr and Braun[9], which contain contributions from all internal variables, i.e. $S_i(T, P, C_j, \Phi_k)$. In an inhomogeneous system, the free energy is in principle a function of both internal variables and their gradients, all state functions, including the partial molar entropy are also a function of the gradients. This is in analogy to the generalized gradient approximations in density functional theory [10] and the gradient energy term in free energy functions used in phase-field simulations of internal processes [11], and they can be important when internal variables undergo rapid changes in the system.

It should also be pointed out that the gradients in Eq. 6 include both gradients of potentials and molar quantities with the latter being $\frac{dC_j}{dx}$. Since the Onsager reciprocal relations are already naturally fulfilled by Eq. 1 and for potential gradients only, further discussions on whether the Onsager reciprocal relations are valid or not with the gradients of both potentials and molar quantities are not only unnecessary, but also confusing.

This confusion made Kocherginsky and Gruebele [2] conclude that "It also becomes clear that if two fluxes are carried simultaneously by several species, each with different concentration, mobility, and molar properties, the traditional reciprocal relations are not valid and need to be generalized." Similarly, their statement [2] that "The forces are given by gradients of driving factors such as temperature, pressure, chemical potential, concentration" is also problematic because the chemical potential and concentration of a species are conjugate variables and are not independent of each other. This confusion also resulted in their conclusion[1] that the ratio of the diagonal coefficient in a binary system, i.e. $L_{22}/L_{11}$, equals the square of the molar charge of ion $i$. This is not true in general because the kinetic coefficients of independent species are independent of each other in general [6].